\documentclass[%
 reprint,
 amsmath,amssymb,
 aps,
 pra,
 floatfix
]{revtex4-2}

\usepackage{graphicx}
\usepackage{dcolumn}
\usepackage{bm}

\usepackage{amsmath}
\usepackage{amssymb}
\usepackage{braket}
\usepackage{mathrsfs}
\usepackage{hyperref}
\usepackage{xfrac}
\usepackage{upgreek}

\renewcommand\bra[1]{{\langle{#1}|}}
\makeatletter
\renewcommand\ket[1]{%
  \@ifnextchar\bra{\k@t{#1}\!}{\k@t{#1}}%
}
\newcommand\k@t[1]{{|{#1}\rangle}}
\makeatother

\begin{document}

\preprint{APS/123-QED}

\title{Quantum limits to chiroptical molecular discrimination}
\author{Mikael P. Backlund}
 \email{mikaelb@illinois.edu}
\affiliation{%
 Department of Chemistry,
 Illinois Quantum Information Science and Technology Center (IQUIST), and Center for Biophysics and Quantitative Biology, University of Illinois at Urbana-Champaign, Urbana, IL, USA 61801
}%

\date{\today}

\begin{abstract}
Discriminating enantiomer pairs is of central importance in the molecular sciences. The preferential interaction between chiral light and matter is commonly employed in this discrimination, despite fundamental challenges. Here we present quantitative error bounds for classifying the handedness of a chiral, randomly oriented, quantum optical emitter from the perspective of classical and quantum hypothesis testing. Most interestingly, we find that a collective measurement on $N$ photons can provide an orders-of-magnitude improvement in error rate relative to a corresponding separable measurement.
\end{abstract}

\maketitle


\section{Introduction}

Preferential interactions between circularly polarized light and chiral molecules have fascinated scientists for well over a century \cite{barron2004molecular,berova2012comprehensive}. Harnessing chiroptical emission and absorption is a major goal in the development of numerous promising technologies \cite{vanorman2025chiral}. Biology is exquisitely sensitive to chirality \cite{blackmond2010origin}, and accurate enantiomeric identification and purification can be a matter of life or death in medicine \cite{tokunaga2018understanding,nguyen2006chiral}. Chiroptical spectroscopies comprise some of the most widely employed analytical techniques for distinguishing a molecule from its mirror-image counterpart, despite being fundamentally difficult due to the mismatch between optical wavelengths and molecular dimensions \cite{craig1998molecular}.

Molecules that exhibit differential absorption or emission of left- vs. right-circularly polarized (LCP vs. RCP) light are ubiquitously characterized by the dissymmetry factor:
\begin{equation}
    \mathscr{G} = 2 \frac{I_L - I_R}{I_L + I_R} \in [-2,2],
\end{equation}
where $I_L$ denotes the intensity of LCP light absorbed (in the case of circular dichroism, CD) or emitted (in the case of circularly polarized luminescence, CPL) and $I_R$ is that of RCP \cite{barron2004molecular,berova2012comprehensive}. Even for a hypothetical molecule with an unusually large $|\mathscr{G}| = 0.5$, about 4 of every 10 photons absorbed/emitted will be of the ``wrong'' handedness. Thus the discrimination of pairs of chiral molecules based on the absorption or emission of optical photons is an inherently statistical endeavor. As such, we stand to gain fundamental insights from rigorously analyzing the problem as an exercise in statistical inference.

The application of classical detection and estimation theory \cite{vantrees2013detection,cover2006elements} to problems in chemical metrology has become increasingly widespread, especially in the context of single-molecule microscopy \cite{chao2016fisher}. More recently, our group and others \cite{BacklundPRL2018,MitchellPRA2022,MitchellPRA2023,dingilian2026quantifying,mitchell2026quantum,ZhangPRL2019,zhang2021single,ChenPRL2025} have found utility in analyzing various tasks of molecular microscopy from the vantage of quantum detection and estimation theory \cite{helstrom1976quantum}. In this Letter, we consider the assignment of the handedness of an emitter as a classical and quantum symmetric binary hypothesis test. In doing so, we establish quantitative speed limits for chiroptical discrimination. We confirm that simply counting $L$ vs. $R$ photons is the best-possible classical strategy. Surprisingly, however, we find that an optimal quantum measurement can in principle vastly improve the error probability given a fixed photon budget.

\begin{figure}
    \centering
    \includegraphics[width=\linewidth]{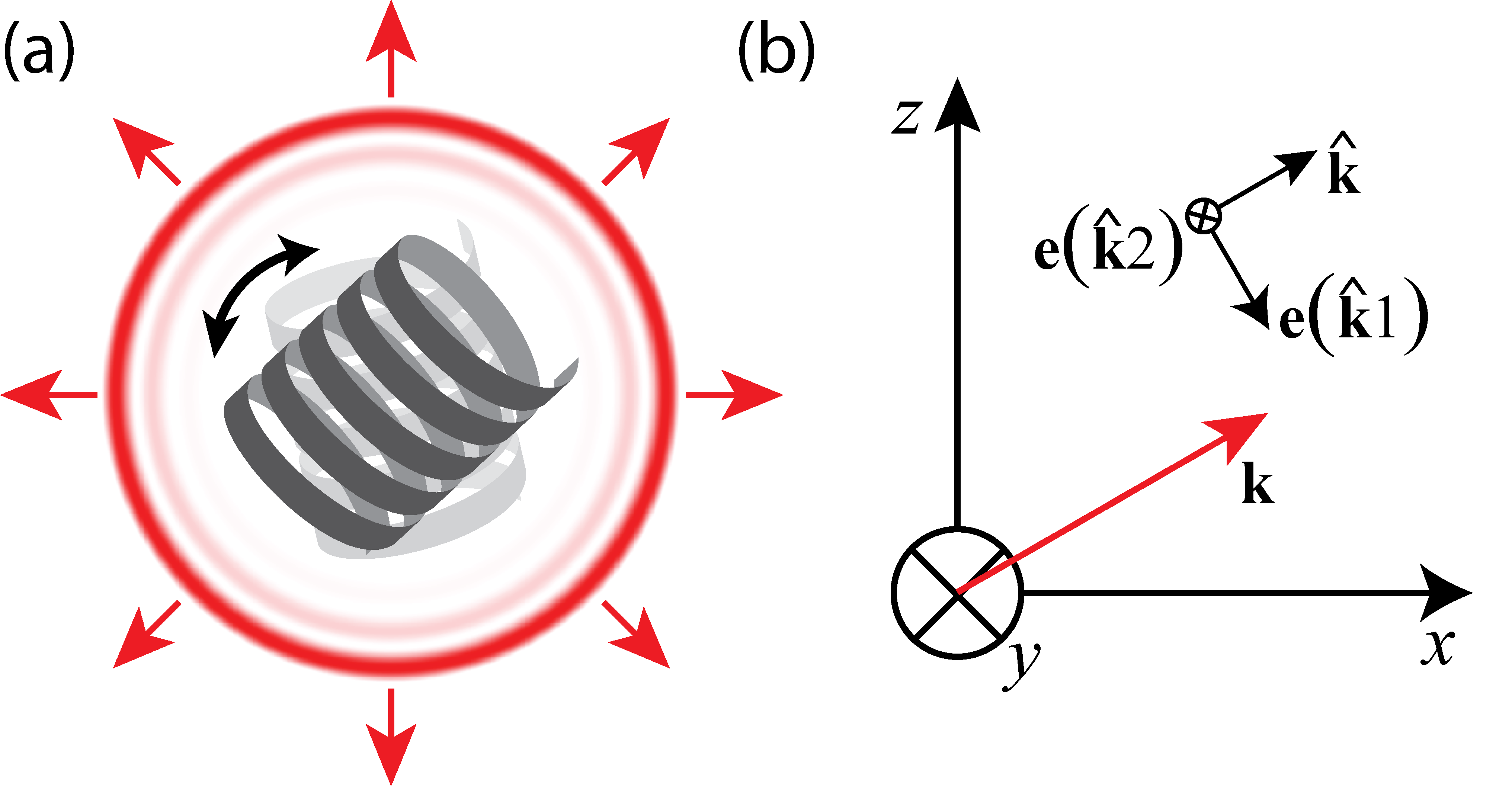}
    \caption{(a) Illustration of the model. We consider the spherically symmetric field due to a randomly oriented chiral quantum emitter, here depicted as a tumbling helix. (b) Lab-frame coordinate definitions, with wavevector $\mathbf{k}$. The unit vectors $\mathbf{e}\left(\hat{\mathbf{k}}1\right)$, $\mathbf{e}\left(\hat{\mathbf{k}}2\right)$, and $\hat{\mathbf{k}}$ form a right-handed triad. For each $\hat{\mathbf{k}}$, the unit circular polarization vectors are $\mathbf{e}\left(\hat{\mathbf{k}}L\right)=\left[\mathbf{e}\left(\hat{\mathbf{k}}1\right)+i\mathbf{e}\left(\hat{\mathbf{k}}2\right)\right]/\sqrt{2}$ and $\mathbf{e}\left(\hat{\mathbf{k}}R\right)=\left[\mathbf{e}\left(\hat{\mathbf{k}}1\right)-i\mathbf{e}\left(\hat{\mathbf{k}}2\right)\right]/\sqrt{2}$.}
    \label{fig:overview}
\end{figure}

Though it is somewhat more niche than CD, we will focus here on CPL for two main reasons. For one, normalizing for the number of photons collected/detected builds in a natural accounting of resources, thus facilitating apples-to-apples comparisons. Second, as we have already established, much of our motivation comes from the world of single-molecule microscopy, where fluorescence detection is standard. Chiroptical microscopy of single molecules was reported two decades ago \cite{hassey2006probing}, but subsequent controversy over polarization artifacts cast doubt on the origins of the anomalously large $\mathscr{G}$ factors observed \cite{tang2009limits,barnes2009comment,cohen2009reply,wakabayashi2014anisotropic,deng2016theoretical}. These events helped birth an area of active research in which enhanced $\mathscr{G}$ factors are engineered using structured excitation light \cite{tang2011enhanced} and/or tailored near-field environments \cite{avalos2022chiral,hentschel2017chiral,kakkanattu2021review,solomon2018enantiospecific,solomon2020nanophotonic,warning2021nanophotonic}. In the current work, we consider the fundamental limits of chiroptical discrimination from a complementary perspective, assuming far-field collection from a randomly oriented quantum emitter with handedness denoted by $+$ or $-$ (Fig. \ref{fig:overview}).

By focusing on emission/collection/detection side of the apparatus, our work is distinguished from previous efforts in pursuit of a quantum enhancement to chiral discrimination, which have tended either to employ quantum excitation light \cite{tischler2016quantum,matsuzaki2024sub} and/or exploit differences in coherent, cyclic population transfer \cite{kral2001cyclic,kral2003two,li2008dynamic,chen2021enantio,chen2022enantiodetection,zou2022enantiodiscrimination,ye2021entanglement,ye2023single}. We note that chiral molecular discrimination was recently also framed as a quantum hypothesis test in Ref. \cite{ye2023single}. 

In a companion paper we derive the following state of the field due to such a source \cite{Backlund2026Article}:
\begin{eqnarray} \label{eq_rhopmusefuldefn}
    \bar{\rho}_\pm = &&\frac{1}{2 \left(\sum_\mathbf{k}k\left|\Phi(k)\right|^2\right)}\sum_{\mathbf{k},\mathbf{k'},s,s'} \Phi(k)\Phi^*(k')\sqrt{k k'} \nonumber \\ &&\times G_{ss'}^{(\pm)}(\mathbf{k},\mathbf{k'})
    \ket{1\left(\mathbf{k}s\right)}\bra{1\left(\mathbf{k'}s'\right)},
\end{eqnarray}
where $\ket{1(\mathbf{k}s)} = \hat{a}^\dagger(\mathbf{k}s)\ket{\text{vac}}$ denotes the one-photon state for the mode with wavevector $\mathbf{k}$ and polarization $s\in\{L,R\}$, and $\Phi(k)$ is a spectral function depending on the wavenumber but not the direction or polarization of the mode. Directional correlations are encoded in the functions
\begin{subequations} \label{eq_Gdefwithepsilons}
    \begin{eqnarray}
        G_{LL}^{(\pm)}(\mathbf{k},\mathbf{k'}) = \left[ \epsilon_\mu  \mp \frac{\mathscr{G}}{2} + \epsilon_m + \tilde{\epsilon}_Q k_j k'_j\right]e_i^*\left(\mathbf{\hat{k}}L\right)e_i\left(\mathbf{\hat{k}'}L\right) \nonumber \\ + \tilde{\epsilon}_Q k_j k'_i e_i^*\left(\mathbf{\hat{k}}L\right)e_j\left(\mathbf{\hat{k}'}L\right) \nonumber \\
    \end{eqnarray}
    \begin{eqnarray}
        G_{RR}^{(\pm)}(\mathbf{k},\mathbf{k'}) = \left[ \epsilon_\mu \pm \frac{\mathscr{G}}{2} + \epsilon_m + \tilde{\epsilon}_Q k_j k'_j\right]e_i^*\left(\mathbf{\hat{k}}R\right)e_i\left(\mathbf{\hat{k}'}R\right) \nonumber \\ + \tilde{\epsilon}_Q k_j k'_ie_i^*\left(\mathbf{\hat{k}}R\right)e_j\left(\mathbf{\hat{k}'}R\right) \nonumber \\
    \end{eqnarray}
    \begin{eqnarray}
        G_{LR}(\mathbf{k},\mathbf{k'}) = \left[ \epsilon_\mu - \epsilon_m + \tilde{\epsilon}_Q k_j k'_j\right]e_i^*\left(\mathbf{\hat{k}}L\right)e_i\left(\mathbf{\hat{k}'}R\right) \nonumber \\ + \tilde{\epsilon}_Q k_j k'_ie_i^*\left(\mathbf{\hat{k}}L\right)e_j\left(\mathbf{\hat{k}'}R\right) \nonumber \\
    \end{eqnarray}
\end{subequations}
with $G_{RL}$ obtained from $G_{LR}$ upon swapping the polarization indices, and where $\mathbf{e}\left(\mathbf{\hat{k}}s\right)$ is the complex unit polarization vector for the corresponding mode. We assume the Einstein summation convention throughout. In Eq. (\ref{eq_Gdefwithepsilons}), $\epsilon_\mu$, $\epsilon_m$, $\mathscr{G}$, and $\tilde{\epsilon}_Q$ are related to the electric dipole $\boldsymbol{\upmu^{01}}\in\mathbb{R}^3$, magnetic dipole $\mathbf{m^{01}}\in\mathbb{I}^3$, and electric quadrupole $\mathbf{Q^{01}}\in\mathbb{R}^{3\times3}$ moments of the molecular transition. If we define the generalized ``dipole'' strength \cite{gendron2019ab} via
\begin{eqnarray}
    D^{01} &&\equiv\left(\mu^{01}\right)^2 + \frac{1}{c^2}\left|m^{01}\right|^2 \nonumber \\  &&+ \frac{1}{10}\left(3\left\lVert\mathbf{Q}^{01}\right\rVert_F^2 - \left(\text{Tr}\,\mathbf{Q^{01}}\right)^2\right)\frac{\sum_\mathbf{k}k^3\left|\Phi(k)\right|^2}{\sum_\mathbf{k}k\left|\Phi(k)\right|^2}, \nonumber \\
\end{eqnarray}
where the Frobenius norm of a matrix $\mathbf{A}$ is defined by $\left\lVert\mathbf{A}\right\rVert_F^2 = \sum_{ij}\left|A_{ij}\right|^2$, then $\epsilon_\mu \equiv \left(\mu^{01}\right)^2/D^{01} $ is the unitless relative electric dipole strength, $\epsilon_m \equiv \left(m^{01}\right)^2/(c^2 D^{01})$ is the unitless relative magnetic dipole strength, the dissymmetry factor can be written \cite{barron2004molecular,berova2012comprehensive}:
\begin{equation}
    \mathscr{G} \equiv \frac{4\text{Im}\left(\boldsymbol{\upmu^{10}}\cdot\mathbf{m^{01}}\right)}{c D^{01}} = 4\sqrt{\epsilon_\mu \epsilon_m}\cos\theta_{\mu m},
\end{equation}
where $\theta_{\mu m}$ is the angle between the transition electric and magnetic dipoles, and the parameter $\tilde{\epsilon}_Q$ has units of [length]$^2$ and is defined:
\begin{equation}
    \tilde{\epsilon}_Q \equiv \frac{3\left\lVert\mathbf{Q}^{01}\right\rVert_F^2 - \left(\text{Tr}\,\mathbf{Q^{01}}\right)^2}{10 \,D^{01}}.
\end{equation}
The unitless relative electric quadrupole strength is:
\begin{equation}
    \epsilon_Q \equiv \tilde{\epsilon}_Q\frac{\sum_\mathbf{k}k^3\left|\Phi(k)\right|^2}{\sum_\mathbf{k}k\left|\Phi(k)\right|^2}.
\end{equation}
or, upon taking the continuum limit $\sum_\mathbf{k} \to \frac{V}{(2\pi)^3}\int\mathrm{d}^3\mathbf{k}$:
\begin{equation}
    \epsilon_Q = \tilde{\epsilon}_Q\frac{\int_0^\infty\mathrm{d}k \, k^5 \left|\Phi(k)\right|^2}{\int_0^\infty\mathrm{d}k \, k^3 \left|\Phi(k)\right|^2} \approx \tilde{\epsilon}_Q k_0^2 \equiv \epsilon_Q^0,
\end{equation}
where the approximate equality on the right holds in the case $\left|\Phi(k)\right|^2 \to \delta(k-k_0)$. The definitions above imply the normalization condition $\epsilon_\mu + \epsilon_m + \epsilon_Q = 1$.

In Ref. \cite{Backlund2026Article} we show that $\bar{\rho}_\pm$ can be conveniently represented by a finite-rank, block-diagonal density matrix of the form:
\begin{equation} \label{eq_rhomatrix}
    \boldsymbol{\bar{\uprho}_\pm} =
        \begin{pmatrix}
  \boldsymbol{\bar{\uprho}_\pm^{(\parallel)}} & \mathbf{0} \\  
  \mathbf{0} & \boldsymbol{\bar{\uprho}^{(\mathscr{\perp})}}. 
\end{pmatrix}
\end{equation}
The submatrix $\boldsymbol{\bar{\uprho}^{(\perp)}}$ is the same for both stereoisomers, and it's trace is the probability of having emitted the photon via the electric quadrupole, i.e.:
\begin{equation}
    \text{Tr}\boldsymbol{\bar{\uprho}^{(\perp)}} = \epsilon_Q.
\end{equation}
The submatrix $\boldsymbol{\bar{\uprho}_\pm^{(\parallel)}}$ is given explicitly by:
\begin{equation}
    \boldsymbol{\bar{\uprho}_\pm^{(\parallel)}} = \left(\epsilon_\mu + \epsilon_m\right)\frac{\mathbf{I_3}}{3} \otimes \varrho_\pm,
\end{equation}
where $\mathbf{I_3}$ is the $3\times3$ identity matrix and we've defined the normalized qubit states:
\begin{equation} \label{eq_qubitstates}
    \varrho_\pm \equiv \frac{1}{2}\left(I + \tilde{\chi}\sigma_x \mp \frac{\tilde{\mathscr{G}}}{2} \sigma_z\right),
\end{equation}
with $\tilde{\chi} \equiv (\epsilon_\mu-\epsilon_m)/(\epsilon_\mu+\epsilon_m) \in [-1,1]$ and \mbox{$\tilde{\mathscr{G}} \equiv \mathscr{G}/(\epsilon_\mu+\epsilon_m)$}.
\section{Results and Discussion}
With $\bar{\rho}_\pm$ in hand, we now consider the task of chiroptical discrimination from the vantage of quantum symmetric binary hypothesis testing \cite{audenaert2007discriminating}. The scenario is that we are given $N$ copies of the photon in state $\bar{\rho}_+$ or $\bar{\rho}_-$ such that the overall quantum state is $\bar{\rho}_+^{\otimes N}$ or $\bar{\rho}_-^{\otimes N}$, with both being equally likely a priori. Our goal is to perform the quantum measurement that allows us to determine the handedness of the emitter with minimal probability of error. By treating the stream of photons as otherwise indistinguishable, we've assumed that $\Phi(k)$ is sufficiently narrow around $k_0$, as realized, for instance, with a narrow-band etalon \cite{deng2019quantum}.

To begin, we take $N=1$ such that the task is to determine whether the underlying state is $\bar{\rho}_+$ or $\bar{\rho}_-$. The minimum probability of error is given by the Helstrom bound \cite{helstrom1976quantum}:
\begin{equation} \label{eq_Helstrombound}
    P_{\text{e,min}}^\text{(H)} = \frac{1}{2}\left( 1 - \frac{1}{2}\text{Tr}\left| \bar{\rho}_+-\bar{\rho}_- \right| \right),
\end{equation}
where $|\mathbf{A}| = \sqrt{\mathbf{A}^\dagger \mathbf{A}}$. Using the matrix form of $\bar{\rho}_\pm$ from Eq. (\ref{eq_rhomatrix}) gives:
\begin{equation} \label{eq_PHelstrom1photon}
    P_\text{e,min}^\text{(H)} = \frac{1}{2}\left( 1 - \frac{|\mathscr{G}|}{2} \right).
\end{equation}

The conventional approach to measuring CPL is to split the collected light into $L$ and $R$ polarizations and count photons. For the one-photon measurement considered in this section this is equivalent to noting whether a click is recorded on the $L$ or $R$ detector. It is uncommon to collect light from all $4\pi$ steradians around the emitter, but since collecting more light is always better, we will consider the limiting case in which no $\mathbf{k}$ is lost, e.g. resulting from collection with opposing objective lenses \cite{hell1992properties} in the limit of very high numerical aperture (NA). The relevant measurement then is given by $\{E_L,E_R\}$, with positive operator-valued measures (POVMs) \cite{WisemanMilburn2010}:
\begin{equation}
    E_s = \sum_\mathbf{k} \ket{1(\mathbf{k}s)}\bra{1(\mathbf{k}s)}, \quad s\in\{L,R\}.
\end{equation}
The probability of recording the photon in either channel is given by Born's rule:
\begin{equation}
    p_\pm(s) = \text{Tr}\left(E_s \bar{\rho}_\pm\right) = \sum_\mathbf{k} \braket{1(\mathbf{k}s)|\bar{\rho}_\pm|1(\mathbf{k}s)},
\end{equation}
yielding
\begin{equation}
    p_\pm(s) = \frac{1}{2}\left( 1 \mp (-1)^{\delta_{s,R}} \frac{\mathscr{G}}{2} \right).
\end{equation}
Based on these classical probabilities, the minimum probability of error is given by Bayes' error \cite{vantrees2013detection}:
\begin{equation}
    P_\text{e,min}^\text{(B)} = \frac{1}{2}\left( 1 - \frac{|\mathscr{G}|}{2} \right).
\end{equation}
Comparing to Eq. (\ref{eq_PHelstrom1photon}) we conclude that $P_\text{e,min}^\text{(B)}=P_\text{e,min}^\text{(H)}$, and thus that the conventional approach to chiroptical discrimination cannot be beat in the one-photon case. 

For physically relevant $\mathscr{G}$ values, however, the measurement $\{E_L,E_R\}$ must be implemented repeatedly on many photons in order to obtain a tolerable error probability. In the asymptotic limit ($N\to\infty$), the probability of error for the symmetric binary hypothesis improves exponentially \cite{Chernoff1952,cover2006elements}:
\begin{equation}
    P_\text{e,min}^\text{(CC)} \sim e^{-N D_\text{CC}\left(p_+,p_-\right)},
\end{equation}
where the classical Chernoff information is
\begin{equation} \label{eq_classicalchernoffinfodef}
    D_\text{CC}(p_+,p_-) = - \log \min_{\sigma \in [0,1]} \sum_{s\in \{L,R\}} p_+^\sigma(s)p_-^{1-\sigma}(s).
\end{equation}
The sum in Eq. (\ref{eq_classicalchernoffinfodef}) obtains its minimum in this case at $\sigma = \frac{1}{2}$. The Chernoff information can be computed directly to be:
\begin{equation} \label{eq_classicalchernoffinfo}
    D_\text{CC}(p_+,p_-) = -\frac{1}{2}\log\left( 1 - \frac{\mathscr{G}^2}{4} \right) \approx \frac{\mathscr{G}^2}{8},
\end{equation}
where the approximate equality holds when $\mathscr{G}^2 \ll 4$. Now suppose we are able to collect $N$ photons before applying a measurement. The state of the composite system is either $\bar{\rho}_+^{\otimes N}$ or $\bar{\rho}_-^{\otimes N}$. Repeated application of $\{E_L,E_R\}$ is one possible choice for the measurement, but quantum mechanics allows for a much wider array of measurement strategies \cite{WisemanMilburn2010,barnett2009quantum}, which in the context of hypothesis testing can be broadly categorized as either separable or collective \cite{audenaert2007discriminating,nussbaum2009chernoff,calsamiglia2008quantum,hayashi2009discrimination,calsamiglia2010local,higgins2009mixed,higgins2011multiple,conlon2023discriminating,conlon2025attainability}. Repeated application of $\{E_L,E_R\}$ corresponds to a particular choice of the former known as a locally optimal fixed local (LOF) measurement \cite{higgins2011multiple}.

\begin{figure}
    \centering
    \includegraphics[width=\linewidth]{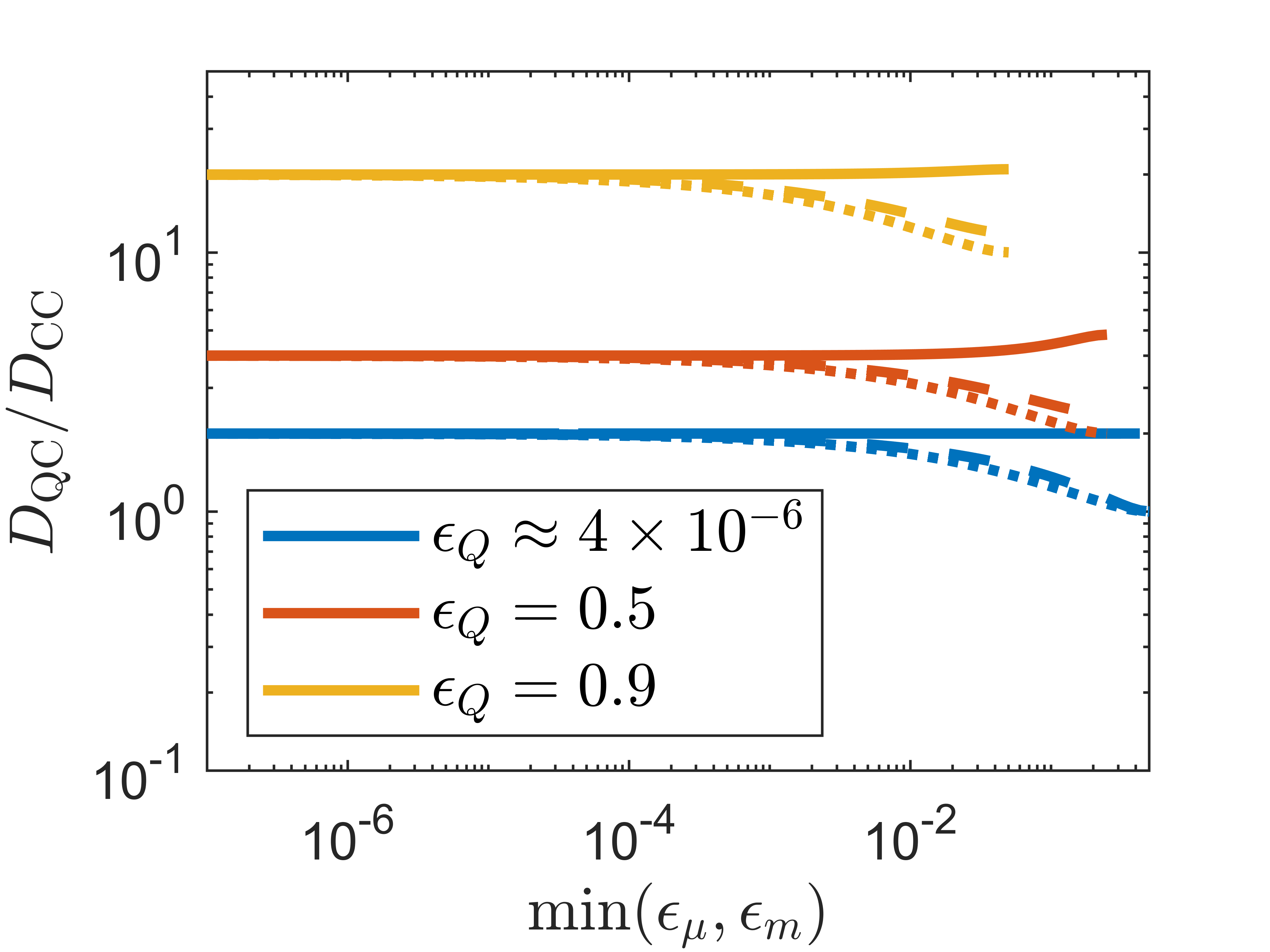}
    \caption{Calculated ratio of quantum to classical Chernoff information for various $\epsilon_\mu$, $\epsilon_m$, $\epsilon_Q$, and $\theta_{\mu m}$. Solid lines: $\theta_{\mu m} = 0^\circ$, dashed lines: $\theta_{\mu m} = 45^\circ$, dotted lines: $\theta_{\mu m} \approx 93.8^\circ$. The values $\theta_{\mu m} \approx 93.8^\circ$ and $\epsilon_Q \approx 4\times10^{-6}$ are based on calculations for a small helicene reported in Ref. \cite{wakabayashi2014anisotropic}.}
    \label{fig:Dratios}
\end{figure}

The error rate of the best collective measurement scheme scales exponentially via \cite{audenaert2007discriminating,nussbaum2009chernoff}:
\begin{equation}
    P_\text{e,min}^\text{(QC)} \sim e^{-N D_\text{QC}\left(\bar{\rho}_+,\bar{\rho}_-\right)},
\end{equation}
where the quantum Chernoff information is given by:
\begin{equation}
    D_\text{QC}(\bar{\rho}_+,\bar{\rho}_-) = - \log \min_{\sigma \in [0,1]} \text{Tr} \left(\bar{\rho}_+^\sigma \bar{\rho}_-^{1-\sigma}\right).
\end{equation}
The form of $\bar{\rho}_\pm$ given by Eqs. (\ref{eq_rhomatrix}-\ref{eq_qubitstates}) leads to the simplified relation:
\begin{equation}
    \text{Tr}\left(\bar{\rho}_+^\sigma \bar{\rho}_-^{1-\sigma}\right) = \left(\epsilon_\mu + \epsilon_m\right)\text{Tr}\left(\varrho_+^\sigma \varrho_-^{1-\sigma}\right) + \epsilon_Q.
\end{equation}
This quantity takes its minimum on the interval at $\sigma = \frac{1}{2}$. A direct calculation yields:
\begin{equation}
    D_\text{QC}(\bar{\rho}_+,\bar{\rho}_-) = -\log\left(1 - \frac{\mathscr{G}^2}{4}\Upsilon\right),
\end{equation}
where
\begin{equation}
    \Upsilon \equiv \frac{\epsilon_\mu + \epsilon_m - \sqrt{4\epsilon_\mu \epsilon_m - \mathscr{G}^2/4}}{\left(\epsilon_\mu-\epsilon_m\right)^2 + \mathscr{G}^2/4}.
\end{equation}
If either $\epsilon_\mu \gg \epsilon_m,\epsilon_Q$ (as is typically the case for small organic chiroptical emitters \cite{sanchez2015circularly,zhao2019advances,mori2021chiroptical}) or $\epsilon_m \gg \epsilon_\mu,\epsilon_Q$ (as is typically the case for lanthanide-based chiroptical emitters with large dissymmetry factors \cite{muller2009luminescent,carr2012lanthanide,zinna2015lanthanide}) then $\Upsilon \approx 1$, and if $\mathscr{G}^2 \ll 4$ as well then it follows that
\begin{equation} \label{eq_quantumchernoffinfo}
    D_\text{QC}(\bar{\rho}_+,\bar{\rho}_-) \approx \frac{\mathscr{G}^2}{4}.
\end{equation}
Comparing Eq. (\ref{eq_classicalchernoffinfo}) and Eq. (\ref{eq_quantumchernoffinfo}) we conclude:
\begin{equation}
    \frac{D_\text{QC}\left(\bar{\rho}_+,\bar{\rho}_-\right)}{D_\text{CC}(p_+,p_-)} \approx 2
\end{equation}
in this limit, and thus for a given number of photons there is potential for a quadratic improvement in error probability over that produced by the typical LOF measurement, $\{E_L,E_R\}$. In general, the ratio $D_\text{QC}\left(\bar{\rho}_+,\bar{\rho}_-\right)/D_\text{CC}(p_+,p_-)$ depends on each of $\epsilon_\mu$, $\epsilon_m$, $\epsilon_Q$, and $\theta_{\mu m}$ (Fig. \ref{fig:Dratios}). Even larger enhancements are possible if $\epsilon_Q$ is significant.

To make progress toward an experiment that might realistically recover the predicted quantum advantage, we will next consider post-selecting the collected photon by different arrangements of spatial, spectral, and polarization filters. Much of the prior literature investigating the performance of the best separable measurement schemes relative to quantum bounds has focused on the canonical example of qubit discrimination \cite{calsamiglia2008quantum,calsamiglia2010local,higgins2011multiple,conlon2025attainability}. The filtering schemes considered below allow us to map our task to a suitable discrimination of qubits, which in turn will facilitate direct connections to the existing quantum hypothesis testing literature.

\begin{figure}
    \centering
    \includegraphics[width=\linewidth]{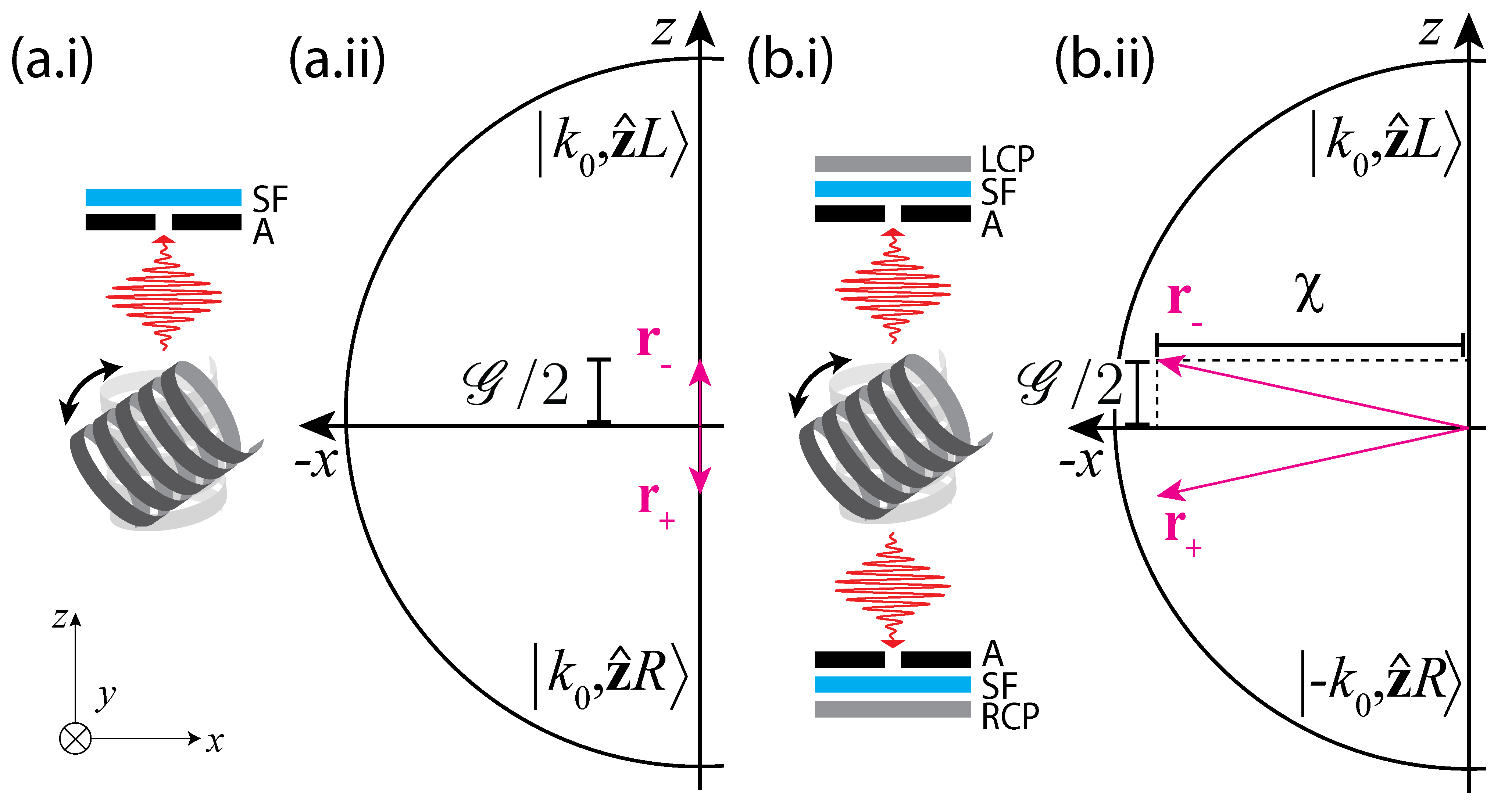}
    \caption{Filtered collection from one (a.i) or two (b.i) directions. A: aperture, SF: spectral filter, LCP/RCP: left/right-circular polarizer. (a.ii) and (b.ii) sketch the Bloch vectors $\mathbf{r}_\pm$ for the qubit states described in the main text, assuming $\chi,\mathscr{G}>0$.}
    \label{fig:collectionschemes}
\end{figure}


The first such arrangement (Fig. \ref{fig:collectionschemes}a.i) gives a simplified version of the one-photon state implicated in the conventional approach to measuring CPL. An aperture and spectral filter can be used to effectively select for a single wavenumber and propagation direction, say $k_0$ and $\mathbf{\hat{z}}$, respectively. In the basis $\left\{\ket{1(k_0\hat{\mathbf{z}},L)},\ket{1(k_0\hat{\mathbf{z}},R)}\right\}$ the so-filtered state of the field is:
\begin{equation}
    \bar{\rho}_\pm^{(\hat{\mathbf{z}})} = \frac{1}{2}\left( I \mp \frac{\mathscr{G}}{2}\sigma_z \right),
\end{equation} 
where $I$ is the identity and $\sigma_z$ is the Pauli-$z$ operator. The relevant Bloch vectors are sketched in Fig. \ref{fig:collectionschemes}a.ii. In Ref. \cite{higgins2011multiple} they consider the discrimination of generic qubits of the form:
\begin{equation} \label{eq_qubitfromHiggins}
    \rho_\pm = \frac{1}{2}\left[I + (1-\nu)\left(\sigma_z \cos\alpha \pm \sigma_x\sin\alpha \right) \right],
\end{equation}
where $\nu \in [0,1]$ quantifies the degree of mixedness. Up to an unimportant rotation we see that $\bar{\rho}_\pm^{(\hat{\mathbf{z}})}$ assumes this form with $\alpha = \pi/2$ and $\nu = 1-|\mathscr{G}|/2$.

In terms of this parameterization, Eq. (17) of Ref. \cite{higgins2011multiple} gives the quantum Chernoff information:
\begin{equation} \label{eq_quantumChernoffqubit}
    D_\text{QC}(\rho_+,\rho_-) = -\log\left[ 1 - (1-c^2)\left(1-\sqrt{1-(1-\nu)^2}\right) \right],
\end{equation}
where $c = \cos\alpha$. Equation (25) of Ref. \cite{higgins2011multiple} gives the classical Chernoff information associated with repeated implementation of the LOF measurement:
\begin{equation} \label{eq_classicalChernoffqubit}
    D_\text{CC}^\text{LOF}(\rho_+,\rho_-) = -\frac{1}{2} \log\left[ 1 - \left(1-\nu\right)^2\left(1-c^2\right) \right].
\end{equation}
A direct calculation shows that for our first example
\begin{eqnarray}
    D_\text{QC}\left(\bar{\rho}_+^{(\hat{\mathbf{z}})},\bar{\rho}_-^{(\hat{\mathbf{z}})}\right) &=& D_\text{CC}^\text{LOF}\left(\bar{\rho}_+^{(\hat{\mathbf{z}})},\bar{\rho}_-^{(\hat{\mathbf{z}})}\right) \nonumber \\ &=& -\frac{1}{2}\log\left(1-\frac{\mathscr{G}^2}{4}\right) \nonumber \\ &\approx& \frac{\mathscr{G}^2}{8} \quad \text{if} \quad \mathscr{G}^2\ll4,
\end{eqnarray}
and that this is realized by the obvious measurement $\left\{\Pi_L^{^{(+\hat{\mathbf{z}})}},\Pi_R^{^{(+\hat{\mathbf{z}})}}\right\}$ where $\Pi_s^{^{(+\hat{\mathbf{z}})}}$ is the projection onto $\ket{1(k_0,+\hat{\mathbf{z}},s)}$. Thus by considering only the light emitted in a single direction we have failed to capture the capacity for a quantum enhancement.

In our second example we consider collecting simultaneously from opposite sides ($+\mathbf{\hat{z}}$ and $-\mathbf{\hat{z}}$) of the sample, as in Fig. \ref{fig:collectionschemes}b.i. The scenario can be further simplified by inserting a polarizer set to pass LCP light in the $+\mathbf{\hat{z}}$ channel and another set to pass RCP light in the $-\mathbf{\hat{z}}$ channel. In the basis $\{\ket{1(k_0\hat{\mathbf{z}},L)},\ket{1(-k_0\hat{\mathbf{z}},R)}\}$, the resulting qubit state is:
\begin{equation}
    \bar{\rho}_\pm^{(\mathbf{\hat{z}}L,-\mathbf{\hat{z}}R)} = \frac{1}{2}\left(I - \chi\sigma_x \mp \frac{\mathscr{G}}{2} \sigma_z\right),
\end{equation}
where $\sigma_x$ is the Pauli-$x$ operator and we've introduced
\begin{equation}
    \chi \equiv \epsilon_\mu-\epsilon_m-\epsilon_Q^0\in[-1,1].
\end{equation}
Bloch vector representations of $\bar{\rho}_+^{(\mathbf{\hat{z}}L,-\mathbf{\hat{z}}R)}$ and $\bar{\rho}_-^{(\mathbf{\hat{z}}L,-\mathbf{\hat{z}}R)}$ are sketched in Fig. \ref{fig:collectionschemes}b.ii. Up to an unimportant rotation, $\bar{\rho}_\pm^{(\mathbf{\hat{z}}L,-\mathbf{\hat{z}}R)}$ takes the form prescribed in Eq. (\ref{eq_qubitfromHiggins}) with $\nu = 1-\sqrt{\chi^2 + \mathscr{G}^2/4}$ and $\alpha = \arctan[-\mathscr{G}/(2\chi)]$. The classical Chernoff information associated with the LOF measurement can be computed from Eq. (\ref{eq_classicalChernoffqubit}) to give:
\begin{eqnarray}
    D_\text{CC}^\text{LOF}\left(\bar{\rho}_+^{(\mathbf{\hat{z}}L,-\mathbf{\hat{z}}R)},\bar{\rho}_-^{(\mathbf{\hat{z}}L,-\mathbf{\hat{z}}R)}\right) &=& -\frac{1}{2}\log\left(1-\frac{\mathscr{G}^2}{4}\right) \nonumber \\ &\approx& \frac{\mathscr{G}^2}{8} \quad \text{if} \quad \mathscr{G}^2\ll4, \nonumber \\
\end{eqnarray}
which is saturated by the polarization-resolved von Neumann measurement $\left\{\Pi_L^{^{(+\hat{\mathbf{z}})}},\Pi_R^{^{(-\hat{\mathbf{z}})}}\right\}$.
The quantum Chernoff information can be computed from Eq. (\ref{eq_quantumChernoffqubit}) to give:
\begin{equation} \label{eq_DQCoppoqubit}
    D_\text{QC}\left(\bar{\rho}_+^{(\mathbf{\hat{z}}L,-\mathbf{\hat{z}}R)},\bar{\rho}_-^{(\mathbf{\hat{z}}L,-\mathbf{\hat{z}}R)}\right) = -\log\left(1 - \frac{\mathscr{G}^2}{4}\Upsilon'\right) \approx \frac{\mathscr{G}^2}{4},
\end{equation}
where
\begin{equation}
    \Upsilon' \equiv \frac{1-\sqrt{1-\chi^2-\mathscr{G}^2/4}}{\chi^2 + \mathscr{G}^2/4}.
\end{equation}
The approximate equality in Eq. (\ref{eq_DQCoppoqubit}) holds in the physically relevant scenario that $\mathscr{G}^2\ll4$ and $|\chi|$ is very near 1, in which case:
\begin{equation}
    \frac{ D_\text{QC}\left(\bar{\rho}_+^{(\mathbf{\hat{z}}L,-\mathbf{\hat{z}}R)},\bar{\rho}_-^{(\mathbf{\hat{z}}L,-\mathbf{\hat{z}}R)}\right)}{ D_\text{CC}^\text{LOF}\left(\bar{\rho}_+^{(\mathbf{\hat{z}}L,-\mathbf{\hat{z}}R)},\bar{\rho}_-^{(\mathbf{\hat{z}}L,-\mathbf{\hat{z}}R)}\right)} \approx 2
\end{equation}
and we see that the possibility of a quadratic improvement has been restored. The crux of this effect is that there is a near-unit coherence between modes of opposite $\hat{\mathbf{k}}$ and opposite polarizations, and that this coherence can be exploited to obtain a quantum advantage when measuring multiple copies, at least for large $N$. 

\begin{figure}
    \centering
    \includegraphics[width=\linewidth]{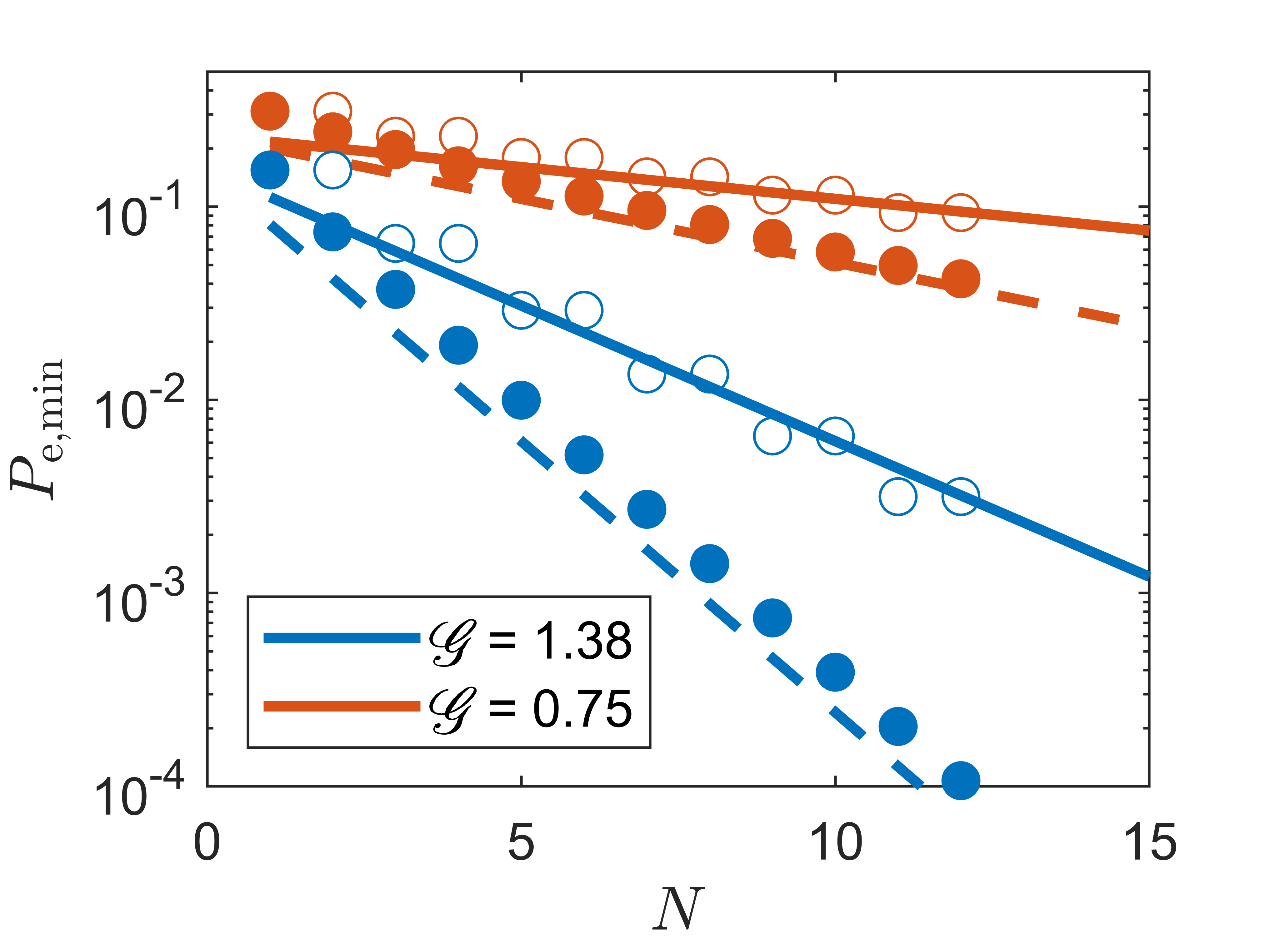}
    \caption{Calculated Bayes' (open circles) and Helstrom (filled circles) error rates for two values of $|\mathscr{G}|$ on the high end of those reported experimentally for lanthanide-based chiroptical emitters. Solid (dashed) lines indicate the scaling of the classical (quantum) Chernoff bounds, with offsets arbitrary.}
    \label{fig:Helstromcalcs}
\end{figure}


Figure \ref{fig:Helstromcalcs} displays numerically computed Helstrom and Bayes error rates associated with discriminating $[\bar{\rho}_+^{(\hat{\mathbf{z}}L,\hat{\mathbf{z}}R)}]^{\otimes N}$ from $[\bar{\rho}_-^{(\hat{\mathbf{z}}L,\hat{\mathbf{z}}R)}]^{\otimes N}$ for various $N\leq 12$. We show data for two values of dissymmetry factors on the high end of what has been reported experimentally for lanthanide complexes exhibiting CPL \cite{petoud2007brilliant,lunkley2008extraordinary}. We've assumed $\epsilon_m > \epsilon_\mu \gg \epsilon_Q^0$ and that the electric and magnetic dipole moments are approximately parallel. Agreement with the classical (solid lines) and quantum (dashed lines) Chernoff bounds indicates rapid convergence to the asymptotic limit. For $\mathscr{G} = 1.38$, the error rate of the optimal classical measurement is over an order of magnitude worse than that of the optimal quantum measurement already by $N = 10$. By extrapolating we can infer that the same milestone would be reached for $\mathscr{G}=0.75$ by around $N=30$. Of course, it becomes difficult to directly compute the Helstrom error for larger $N$ due to the exponential growth in the size of the matrices to be diagonalized.
\section{Conclusion}
These data suggest that a significant improvement in the error rate could feasibly be realized in an experiment focusing on chiral lanthanide complexes possessing large dissymmetry factors. The boost could be enough to make a virtually error-free decision based on measurement of a single, very dim emitter. The outlook is somewhat different for organic chiroptical emitters with much smaller dissymmetry factors ($\sim10^{-3}$), for which a similarly significant improvement in error rate would likely require $\sim$thousands of photons. It is difficult to imagine a scenario in which a collective measurement on 1000 photons would be more practical than a separable measurement on 2000 photons, at least at the current state of technology. Therefore, for organic chiroptical emitters, a reasonable conclusion to draw from our analysis is that even the most sophisticated quantum measurement scheme can only help you so much-- when plagued by small $\mathscr{G}$, directly altering the near field of the emitter may be the only sensible approach to improve chiroptical discrimination \cite{avalos2022chiral,hentschel2017chiral,kakkanattu2021review,solomon2018enantiospecific,solomon2020nanophotonic,warning2021nanophotonic}.

An experimental demonstration on lanthanide-based chiroptical emitters would require hardware at the bleeding edge of current quantum technology. As photon arrival times would be random and infrequent, the information contained therein would first need to be coherently transferred to a suitable quantum memory \cite{thomas2024deterministic,katz2018light,tang2015storage,bhaskar2020experimental,usmani2010mapping}. Only after sufficiently many photons have been collected and buffered are the qubits to be coherently processed and read out. A recent proposal to perform quantum-enhanced imaging of weak optical sources raises similar considerations \cite{mokeev2026enhancing}. Silicon-vacancy centers embedded in diamond nanophotonic cavities seem to be an especially promising platform for physically realizing the required qubit registers \cite{bhaskar2020experimental,stas2022robust,knaut2024entanglement,stas2026entanglement}.

\begin{acknowledgements}
    This work was supported by National Science Foundation award number 2441430. Thanks to Chris Anderson for illuminating discussion.
\end{acknowledgements}

\bibliography{mybib}

\end{document}